\documentclass{article}

\usepackage{arxiv}

\usepackage[utf8]{inputenc} 
\usepackage[T1]{fontenc}    
\usepackage{hyperref}       
\usepackage{url}            
\usepackage{booktabs}       
\usepackage{amsfonts}       
\usepackage{nicefrac}       
\usepackage{microtype}      
\usepackage{lipsum}
\usepackage{graphicx}
\graphicspath{ {./images/} }
\usepackage{listings}

\usepackage{xcolor}

\newcommand{\ag}{\textcolor{black}} 
\newcommand{\ms}{\textcolor{black}}

\title{Structural Chart of Copper-Silver Nanoalloys through machine learning}

\author{
 Manoj Settem \\
  Dipartimento di Ingegneria Meccanica e Aerospaziale\\
  Sapienza Università di Roma\\
  via Eudossiana 18, Roma, Italy, 00184 \\
  \texttt{manoj.settem@uniroma1.it} \\
  \And
Emanuele Telari \\
  Dipartimento di Ingegneria Meccanica e Aerospaziale\\
  Sapienza Università di Roma\\
  via Eudossiana 18, Roma, Italy, 00184 \\
  Departament de Ciència de Materials \\
  i Química Física \& Institut de Química Teòrica i Computacional (IQTCUB) \\
  Universitat de Barcelona\\
  Barcelona, Spain, 08028\\
  \And
Antonio Tinti \\
  Dipartimento di Ingegneria Meccanica e Aerospaziale\\
  Sapienza Università di Roma\\
  via Eudossiana 18, Roma, Italy, 00184 \\
  Laboratory of Molecular Simulation (LSMO) \\
  Institut des Sciences et Ingénierie Chimiques, Ecole
Polytechnique Fédérale de Lausanne (EPFL) \\
Sion 1950, Switzerland\\
  \And
Riccardo Ferrando \\
  Dipartimento di Fisica\\
  Università di Genova\\
  Genova, Italy, 16146 \\
  \texttt{riccardo.ferrando@unige.it} \\
  \And
Alberto Giacomello \\
  Dipartimento di Ingegneria Meccanica e Aerospaziale\\
  Sapienza Università di Roma\\
  via Eudossiana 18, Roma, Italy, 00184 \\
  \texttt{alberto.giacomello@uniroma1.it} \\
}

\begin{document}

\maketitle

\begin{abstract}
Nanoalloys (or alloy nanoparticles) are an important class of materials that are promising for their functional properties. However, designing synthesis protocols to control their structure and chemical ordering is rather challenging. Part of this difficulty stems from the lack of information on their metastable and stable structures. Here, we develop a general computational framework to construct a structural chart of nanoalloys using 38-atom AgCu nanoalloys as a model system. Initially, the equilibrium structural distribution is sampled using parallel tempering combined with molecular dynamics (PTMD). Using a machine learning (ML) based approach, the vast number of sampled configurations are classified into various structural classes. This ML approach produces a single three-dimensional map in which all structures and compositions can be visualized and discriminated. Finally, a finite-temperature structural chart is constructed which provides information on the dominant structures across the entire range of compositions and temperatures. In addition, the structural chart reveals significant differences in thermal stability between nanoalloys and bulk alloys. The presented framework provides an effective route to compute and map the vast structural and chemical space of multicomponent nanoparticles, paving the way to the rational design of functional nanoalloys. 
\end{abstract}

\section{Introduction}
Nanoalloys are nanoparticles consisting of two or more components. In contrast to their single-component counterparts, due to synergistic interactions of their constituents, nanoalloys exhibit interesting catalytic \cite{wu2017CatAgCu,yi2017CatAuNi,sytwuRev2019photoCat,csCatalystsRev}, optical \cite{gaudry2003OptAgCoAgniAuNi,she2012OptAuNi,mueller2020OptSERS}, and magnetic \cite{becerra2015OptAuCo,garfinkel2020MagOpt,boldman2021MagOpt,johnny2021AuCoMag} properties. Depending on the alloyed elements, nanoalloys exhibit a wide variety of chemical orderings, ranging from mixed to separated \cite{FerrChemRev2008}. In the case of alloys that exhibit immiscibility or weak mixing in the bulk, 
it was proposed that miscibility increases at the nanoscale \cite{Christensen1995jpcm}. However, experiments on nanoalloys that are immiscible in the bulk report a wide variety of chemical orderings from complete mixing (solid solution) to separation (core-shell or Janus arrangements) and even inverse core-shell \cite{schnedlitz2019,snellman2021,bogatyrenko2023,sato2017,fan2023}. These results do not provide any resolution on whether the miscibility improves at the nanoscale, as it is unclear whether they are equilibrium or kinetically trapped metastable configurations. 

In order to better understand chemical ordering in nanoalloys, it is desirable to have an equilibrium structural distribution (similar to bulk phase diagrams) that captures the most probable structure as a function of composition and temperature. However, this task is not trivial due to the added complexity of surface effects, nanoalloy size (number of atoms), and non-crystalline structural motifs (icosahedra and decahedra) that become favorable in nanoscale clusters. This is tackled using two different approaches: (1) adapt the thermodynamic modelling used to calculate bulk phase diagrams to nanoparticles by introducing approximations relevant to account for surface and finite-size effects \cite{garzel2012,jabbareh2018}, and (2) statistical mechanics based sampling 
based on (classical) interatomic potentials \cite{panizon2015,wang2012,pohl2012}.

In the first approach, the CALPHAD method is adapted to nanoparticles by adding a Gibbs free energy term corresponding to the surface. This involves accounting for the shape and size of the nanoparticles. As an example, the phase diagram for Ag-Cu nanoalloys has been calculated down to 2 nm \cite{garzel2012,jabbareh2018}. Typically, these phase diagrams reproduce the main characteristics of the bulk phase diagram shifting to lower eutectic temperatures (the eutectic composition moves to lower Cu atom fractions) and relatively enhanced miscibility as the size of nanoparticles decreases. An inherent limitation of this approach is that it does not account for the various structural motifs that are observed only in nanoalloys due to the breaking of translational symmetry, specifically the icosahedral motif which is prevalent in Ag-Cu nanoalloys. In this structure, the atoms close to the center of the particle significantly deviate from the fcc arrangement typical of the bulk alloy. In addition, it is known that close-to-perfect core-shell chemical ordering improves drastically the thermal stability of nanoalloys \cite{rossi2004}. This is in contrast to the decreasing thermal stability near the eutectic composition observed in the calculated nanoalloy phase diagrams.

In the second approach, part of the structural phase diagram (the most probable structure at a given composition and temperature) is typically calculated by considering the detailed atomic arrangement of nanoalloys. For example, the low-temperature region of Pt-Pd nanoalloys (32 to 38 atoms) was calculated by sampling energy minima up to a certain cut-off above the global minimum and estimating the probability of various structural motifs using the harmonic superposition approximation (HSA) \cite{panizon2015}. This allowed the description of the various solid-solid transitions that occur in these nanoalloys. In other instances, Au-Pt \cite{wang2012} and Pt-Rh \cite{pohl2012} nanoalloys were studied using Monte Carlo exploration of octahedral motifs in the region below melting. In the case of Au-Pt (586-atom truncated octahedron), with increasing Pt content, the chemical ordering changes from an Au-rich solid solution to onion-like (where the Pt atoms reside in the subsurface positions) to Janus arrangement to surface segregation of Pt (when Au is insufficient to fully cover Pt). The study of Pt-Rh nanoalloys with truncated octahedra (consisting of 807, 2075, and 9201 atoms) revealed that the nanoscale size increases the stability of ordered regions. Again, there are limitations to these methods: even if all possible structural motifs were sampled with the HSA method, the harmonic approximation becomes inaccurate at increasing temperatures. 

Here, we adopt an enhanced sampling technique, parallel tempering combined with molecular dynamics (PTMD), to sample configurations at all compositions and temperatures (including beyond melting). This was previously applied to mono-metallic nanoparticles \cite{settem2022Au,settem2023AgCuAu}. Here, we apply this method to 38-atom Ag-Cu nanoalloys. There are two main steps in constructing the structural chart: (1) sampling equilibrium configurations using PTMD and (2) classifying the obtained configurations. 
\ag{Because of the wealth of structures and the lack of general classification methods for nanoalloys}, for the second step, we rely on a general machine learning approach for structural classification.
\ms{Prior machine learning (ML) approaches to classification are based on either local atomic coordination arrangement (common neighbour analysis (CNA) \cite{cna1994}) or general structural features as descriptors \cite{parker2020classification,roncaglia2023classification, rapetti2023classification}. However, these approaches have limitations in capturing the global atomic arrangement with respect to the overall particle geometry. We have recently used the radial distribution function (RDF) as the descriptor to classify pure metal nanoparticles \cite{telari2023charting}. RDF has the advantage of capturing the local atomic arrangement as well as the global shape of the nanoparticle.} 
In this work, we adapt the approach of Refs.~\cite{telari2023charting,telari2025isv} to nanoalloys, by  calculating the RDF of atomistic configurations obtained by PTMD at any temperature, feeding them to a convolutional neural network that creates a three-dimensional representation of the configurations, representing it in terms of its relaxed configuration (free from thermal noise or at 0K). This general low-dimensional space is referred to as inherent structure variable (ISV) space \cite{telari2025isv}. 

The accuracy of the equilibrium structural distribution estimated by PTMD depends on the choice of the interatomic potential. Here, we use the Gupta potential to model the various atomic pair interactions. A key characteristic of the Ag-Cu system is the significant lattice mismatch which leads to surface reconstruction in the bulk and nanoalloys. This potential correctly predicts the experimentally observed c(10x2) reconstruction \cite{sprunger1996pHexExpAgCu} of Ag monolayer on bulk Cu \{100\} surfaces \cite{settem2022coreShell}. Due to the size mismatch and the differences in cohesive energy, Ag-Cu nanoalloys exhibit icosahedral motifs in addition to fcc \cite{RapalloJCP2005}. The various icosahedral motifs and surface reconstructions in Ag-Cu nanoalloys predicted by the Gupta potential have been verified against density functional theory (DFT) calculations \cite{settem2022coreShell, bochicchio2010chiralIh, panizon2014DFT}. 

In this work, we analyze how the structure of 38-atoms silver-copper nanoalloys changes across temperatures and compositions. Section~\ref{sec:methods} describes the methods used to sample equilibrium configurations, the machine learning model that maps them to the 3-dimensional ISV space, and HSA calculations. Section~\ref{sec:results} describes the results of the PTMD simulations: the classification and interpretation of the structural motifs of this nanoalloy, their dependence on temperature, and structural transformations. Section~\ref{sec:conclusions} is left for conclusions.

\section{Simulation Approach}
\label{sec:methods}

\begin{figure}[!b]
\centering
  \includegraphics[width=0.8\textwidth]{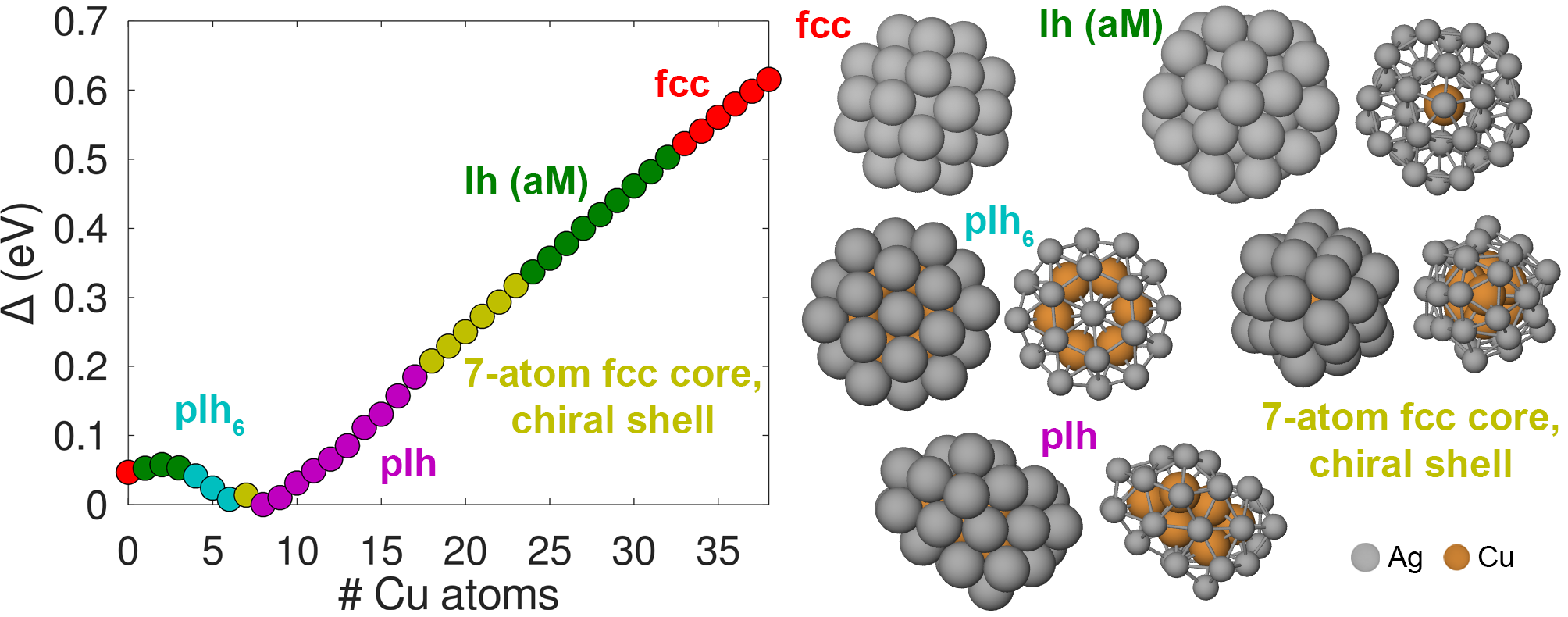}
  \caption{Excess energy ($\Delta$) of global minima. The data points are color-coded according to their structure type as better described in the main text.  Representative structures are reported on the right hand side, with the following compositions: fcc (Ag$_{38}$), Ih (aM) (Ag$_{37}$Cu$_{1}$), pIh$_6$ (Ag$_{32}$Cu$_{6}$), 7-atom fcc core, chiral shell (Ag$_{31}$Cu$_{7}$), and pIh (Ag$_{30}$Cu$_{8}$).}
  \label{fgr:gm_excess_energy_structs}
\end{figure}

\ms{In this work, the main objective is to develop a computational framework to estimate a finite-temperature structural chart of nanoalloys similar to bulk alloy phase diagrams. A key step in this procedure is sampling the equilibrium structural distribution. We chose 38-atom AgCu nanoalloys as model system to establish this framework. AgCu is representative of non-mixing systems for which there is a wide composition range in the bulk with phase separation \cite{pdCuAg}. In addition, AgCu nanoalloys are sluggish with respect to phase separation compared to similar systems like AgNi, AgCo, AuCo \cite{nelli2019,nelli2023freezing}. The choice of 38-atom nanoalloys is to overcome the sluggishness of AgCu system and enable efficient sampling of the equilibrium structural distribution. Also, 38-atom is an ideal size for a regular truncated octahedron which is the global minimum for pure Ag and Cu nanoclusters.} 

We model interactions between Ag and Cu atoms using the Gupta potential \cite{gupta1981,rgl1989}; the parameters are taken from Ref.~\cite{potCuAgGupta}. Initially, we searched for the global minimum of Ag\textsubscript{38-n}Cu\textsubscript{n} nanoalloys at each composition using Monte Carlo combined with basin hopping, an approach referred to as basin hopping Monte Carlo (BHMC) \cite{rossi2009BHMC}, with at least 10\textsuperscript{5} basin hopping steps. This provides the low-temperature picture of how the structural motifs vary with composition. Figure~\ref{fgr:gm_excess_energy_structs} shows the plot of excess energy ($\Delta$) defined as
\begin{equation}
    \Delta=\frac{E-(38-n)\epsilon_{Ag}-n\epsilon_{Cu}}{38^{2/3}} \, ,
\end{equation}
where $E$ is the potential energy of the nanoalloy, $\epsilon_{Ag}$ and $\epsilon_{Cu}$ are the bulk cohesive energies of Ag and Cu, respectively, and $n$ is the number of Cu atoms. We set the value of the lowest excess energy to zero. Low excess energy indicates relatively higher stability. The four lowest excess energy values are found at n\textsubscript{Cu} $=$ 6 to 9 with perfect core-shell chemical ordering (pure Cu core covered by a single layer Ag shell) observed at n\textsubscript{Cu} $=$ 7 and 8. We identified five different structural motifs and the representative structures are shown in Figure \ref{fgr:gm_excess_energy_structs}. Their geometry will be discussed in detail in the Results section.

PTMD simulations are performed for each composition beginning from a geometrically and chemically random configuration. We use 18 replicas in the temperature range 200 K to 1100 K (temperatures are listed in Supplementary Table \ref{sup_tab:replica_T}) and a replica exchange is attempted every 250 ps. After running the PTMD simulations for 250 ns, we run a production phase of an additional 1 {\textmu}s during which configurations are saved at intervals of 125 ps after an exchange attempt. This results in 72,000 configurations per composition and a total of 2,808,000 across all compositions. The PTMD simulations were carried out in LAMMPS \cite{lammps} using a time step of 5 fs.

Once the equilibrium configurations are sampled by PTMD, the next step is to classify them. For this, we adapt a recently developed machine learning approach \cite{telari2023charting, telari2025isv} to Ag\textsubscript{x}Cu\textsubscript{38-x} nanoalloys. The radial distribution function (RDF) of a given nanoparticle configuration (non-minimized with thermal noise included) is first calculated using kernel density estimation \cite{rosenblattKDE, parzenKDE} on the interatomic distances, between 1 {\AA} and 12 {\AA} and discretized into 200 bins. The so-computed RDF is then fed into a convolutional autoencoder, which maps it to the corresponding RDF of the minimized (free of thermal noise) configuration. In the first half of the network, the dimensionality of the input is reduced from 200 to 3, generating a reduced space described in terms of the so-called inherent structure variables (ISVs). The second half of the network then decodes this low-dimensional representation, reconstructing the RDF of the minimized configuration. To achieve this, the neural network is trained by minimizing the mean-square error (MSE) between the actual RDFs of the minimized configuration and the one predicted in its output layer. A specific feature of this neural network is that the input and outputs are not the same as in classical autoencoders.  Complete details of the autoencoder are provided in Ref.~\cite{telari2025isv} and in the \textcolor{black}{Supplementary Section \ref{sup_sec:AE}}. 

The ISV values are calculated for all the PTMD configurations and they are classified into ten structural classes which will be discussed in Sec.~\ref{sec:results}.
Based on the classification, we define a finite temperature structural chart which gives the most probable structural class for any combination of composition and temperature. Since the PTMD simulations are carried out for temperatures 200 K and above, in order to gain structural insights into the low-temperature region, we carried out harmonic superposition approximation (HSA) calculations \cite{settem2022Au} at selected compositions. For HSA calculations, unique configurations (defined as belonging to different structural classes and separated by at least 0.005 meV) up to 0.5 eV above the global minimum were used.

\section{Results}
\label{sec:results}

\subsection{Inherent structure variables (ISV) space}
As anticipated, to identify the geometrical structure of the vast number of configurations (2,808,000) sampled from PTMD, we use a recently developed machine learning approach to produce a low-dimensional inherent structure variables (ISV) representation of the nanoparticle. In contrast to previous instances where this method was applied to pure metal clusters \cite{telari2023charting,telari2025isv}, here we study Ag-Cu nanoalloys. For a bimetallic system, the main challenge for the machine learning model is to achieve a clear separation of the various configurations in terms of both composition and structural type. 

\begin{figure}[!h]
\centering
  \includegraphics[width=1.0\textwidth]{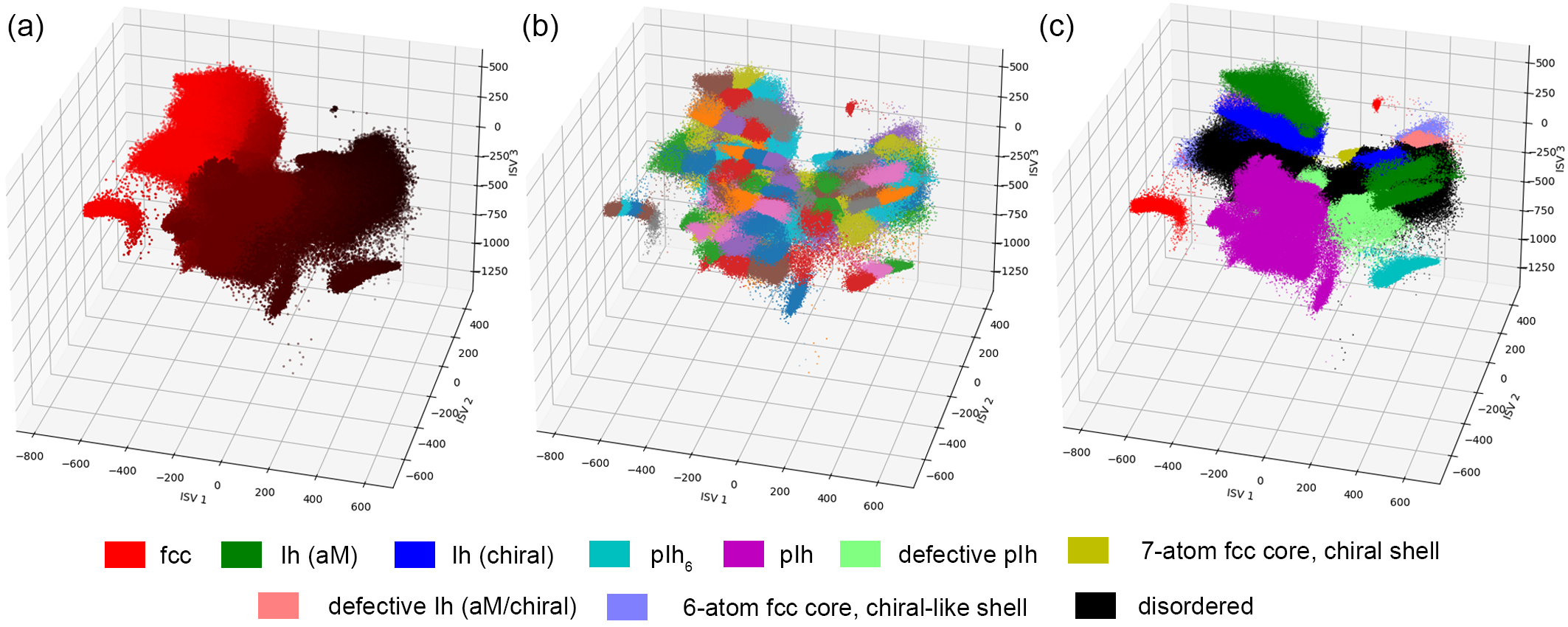}
  \caption{ISV space showing (a) composition map, (b) KMeans clustering with 150 clusters, and (c) final 10 broad structural classes. The color gradient in (a) from black to red indicates alloy composition from pure Cu to pure Ag.}
  \label{fgr:isv}
\end{figure}

We found that using a single RDF (ignoring the atomic species type) mapped onto a 3-dimensional ISV space works best. The composition is encoded in the RDF because the bond lengths between the  atomic species (Ag-Ag, Cu-Cu, and Ag-Cu) are different. \ms{In case of pure metal clusters \cite{telari2025isv}, we found that two ISV variables are sufficient to capture the various structural differences. For nanoalloys, the third ISV variable is needed to account for the composition.} Figure \ref{fgr:isv}a shows the composition map where the configurations are well separated according to composition, which roughly correlates with  ISV-1. The next step is to cluster this space to identify the various structural classes. This step is by far the most challenging due to the multi-modal nature of ISV space. In order to achieve a reasonable classification, we adopt a two-step procedure. First, we over-cluster the space into 150 clusters using KMeans clustering (Figure \ref{fgr:isv}b). Based on visual inspection, we then combine these detailed clusters into ten final structural classes (Figure \ref{fgr:isv}c). The 3D ISV space thus obtained exhibits good structural separation.

\subsection{Structural motifs}

\begin{figure}[!t]
\centering
  \includegraphics[width=0.5\textwidth]{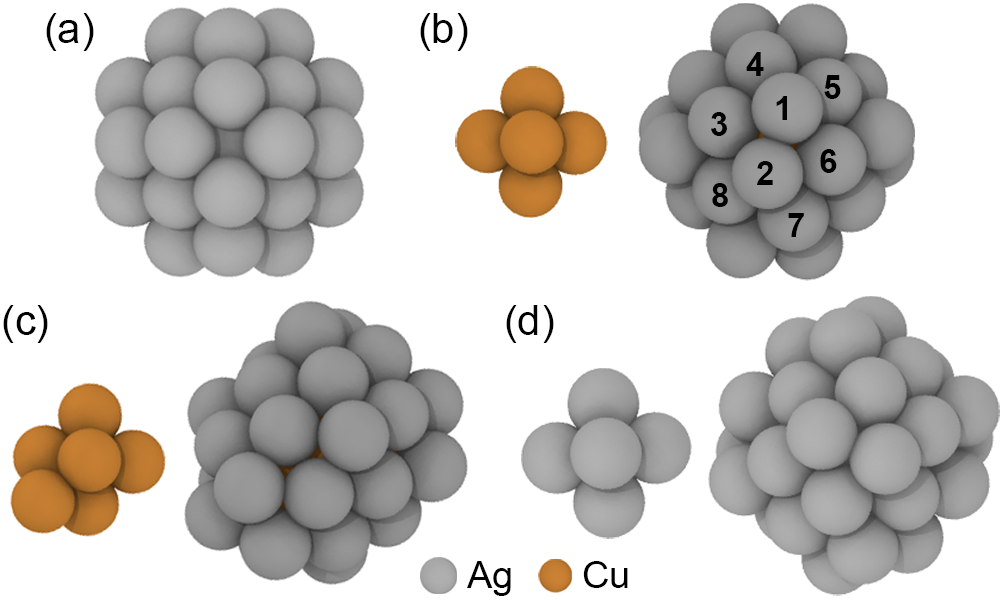}
  \caption{Fcc-based motifs: (a) perfect \emph{fcc}, (b) 6-atom fcc core and the 28-atom chiral shell in a 34-atom \emph{mix} structure, (c) 7-atom fcc core and the 31-atom chiral shell in a 38-atom \emph{mix} structure, (d) 6-atom fcc core with a dense 32-atom shell.}
  \label{fgr:fcc_structs}
\end{figure}

Two broad types of structural motifs were found \ms{in BHMC and PTMD simulations}: fcc and icosahedra; all remaining structural classes are variations thereof. \ms{Figure \ref{fgr:fcc_structs} shows the $fcc$-based motifs which are variations of the 38-atom regular truncated octahedron (Figure \ref{fgr:fcc_structs}a) with minor amounts of defective octahedra consisting of hcp twin planes.} In nanoalloys that exhibit size mismatch (like Ag-Cu), a \emph{mix} structure with the \{111\} facets of the shell in alternating fcc and hcp stacking becomes energetically competitive at small sizes \cite{settem2022coreShell}. A 34-atom \emph{mix} structure (Figure \ref{fgr:fcc_structs}b) consists of 6-atom core and 28-atom chiral shell where the \{111\} facets are rotated resulting in \textbf{T\textsubscript{23}} point group symmetry \ms{in Ag$_{28}$Cu$_6$}. Consequently, the shell attains features resembling the arrangement near an icosahedral vertex (atom groups 1-2-3-4-5-6 and 2-1-6-7-8-3). There are two ways in which 38-atom structures can be obtained from a 34-atom \emph{mix} structure. The first is by increasing the size of the core to 7 atoms (Figure \ref{fgr:fcc_structs}c), which results in a 38-atom \emph{mix} structure. We refer to this structural class as \emph{7-atom fcc core, chiral shell}. This structure has a \textbf{C\textsubscript{3}} point group symmetry \ms{in Ag$_{31}$Cu$_7$ nanoalloys}. On the other hand, a denser shell with four additional atoms also results in a 38-atom structure (Figure \ref{fgr:fcc_structs}d), which has a distorted shell but still retains chiral-like features. We refer to this structural class as \emph{6-atom fcc core, chiral-like shell}.

\begin{figure}[!h]
\centering
  \includegraphics[width=0.5\textwidth]{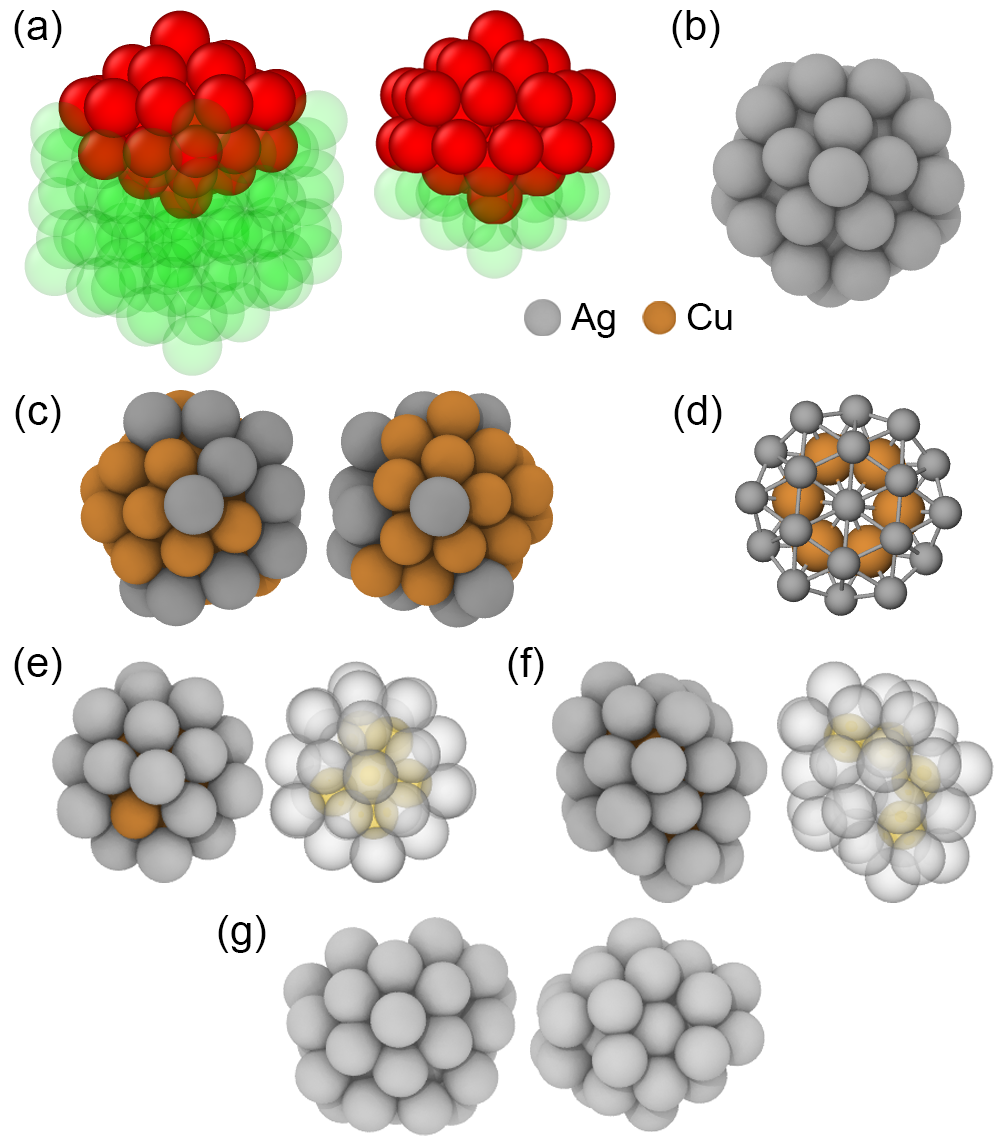}
  \caption{Ih-based motifs. (a) 39-atom structure (in red) carved out from (left) 127-atom anti-Mackay icosahedron and (right) 55-atom Mackay icosahedron. (b) Ih aM (anti-Mackay icosahedron), (c) Ih chiral (chiral icosahedron, shown from both sides), (d) Ag$_{32}$Cu$_{6}$ pIh$_{6}$ (poly-icosahedron with 6 Cu atoms having icosahedral coordination) (e) pIh (poly-Icosahedron, atoms having icosahedral and undefined coordination are shown in yellow and white, respectively) (f) defective pIh (g) defective Ih (aM/chiral).}
  \label{fgr:ih_structs}
\end{figure}

In the case of icosahedra, the basic structure is a 39-atom partial icosahedron with anti-Mackay and Mackay stacking on either side (red structures in Figure \ref{fgr:ih_structs}); this structure can be constructed as a piece of either a 127-atom anti-Mackay icosahedron or a 55-atom Mackay icosahedron (Figure \ref{fgr:ih_structs}a). We define this structure with a missing vertex \emph{Ih (aM)} (anti-Mackay icosahedron) (Figure \ref{fgr:ih_structs}b). When the anti-Mackay and Mackay surfaces rotate, one gets the \emph{Ih (chiral)} structural class  (chiral icosahedron, Figure \ref{fgr:ih_structs}c). The next set of icosahedral structures belongs to the poly icosahedra (pIh) which consist of interpenetrating icosahedra \cite{rossi2004pIh}. For 38-atom clusters, the poly icosahedra are constructed from 13-atom icosahedra. The first structural class is 6-fold poly icosahedra (Figure \ref{fgr:ih_structs}d) consisting of six interpenetrating icosahedra and is referred to as \emph{pIh$_6$} which is highly symmetric for Ag$_{32}$Cu$_6$ exhibiting \textbf{D\textsubscript{6h}} point group symmetry. The other polyicosahedra are less symmetric and are categorized as \emph{pIh}. In addition to these, we also observed defective \emph{pIh} (Figure \ref{fgr:ih_structs}f) and defective \emph{Ih} (aM/chiral, Figure \ref{fgr:ih_structs}g). The defects include partial disordering and a 6-atom ring near the vertices instead of a 5-atom ring (Figure \ref{fgr:ih_structs}g). \ms{Finally, we note that the various structural classes described above refer to the geometrical arrangement irrespective of the chemical arrangement. For example, strictly speaking, the \emph{7-atom fcc core, chiral shell} is truly chiral with \textbf{C\textsubscript{3}} symmetry only at Ag$_{31}$Cu$_7$ composition. However, when other compositions have this same geometrical arrangement, we still assign them to \emph{7-atom fcc core, chiral shell} structural class. The same holds for other structural classes.}

\begin{figure}[!t]
\centering
  \includegraphics[width=1.0\textwidth]{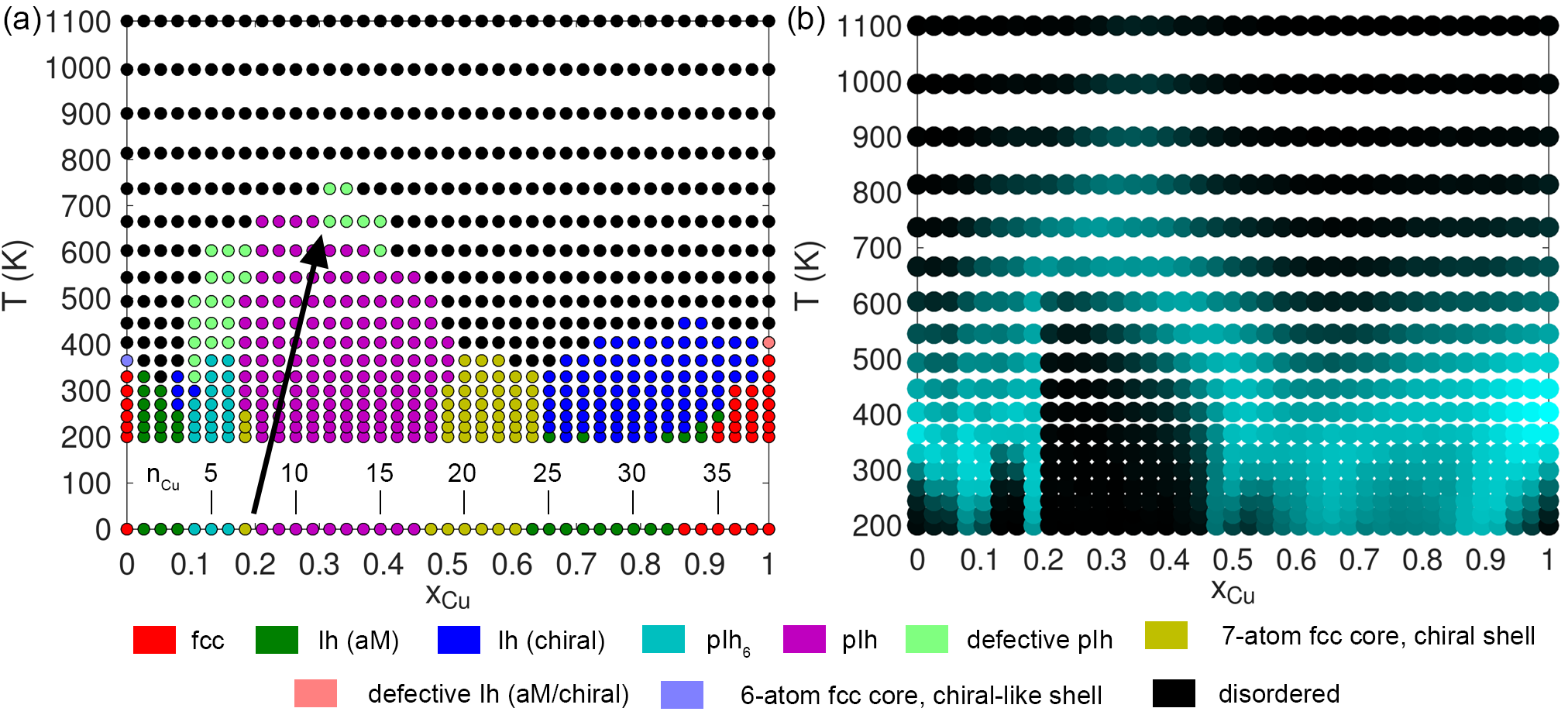}
  \caption{(a) Structural chart of Ag$_{38-n}$Cu$_{n}$ nanoalloys as a function of composition and temperature. \ag{The same structural classes identified in Figure~\ref{fgr:isv} are used.} \ms{The black arrow points from the perfect core-shell arrangement to the composition where the highest thermal stability is observed. (b) Plot of Shannon entropy (the color gradient from black to cyan corresponds to values in the range 0 to 1.76).} }
  \label{fgr:phase_diagram}
\end{figure}

\subsection{Finite temperature structural chart}

Using the structural classification in the ISV-space (Figure \ref{fgr:isv}c), we constructed the structural chart of 38-atom Ag-Cu nanoalloys in the temperature range 200 K to 1100 K. We use the term structural chart as opposed to phase diagram due to the inherent difficulty of defining what a phase is for finite sized particles. \ms{This difficulty stems from characteristic differences between finite-sized nanoalloys and bulk alloys. In case of bulk AgCu, the equilibrium phases are Ag-rich and Cu-rich fcc solid solutions along with liquid. When it comes to nanoalloys, in addition to crystalline fcc structures, non-crystalline structures such as icosahedron and decahedron, are also observed. In bulk alloys, the extremely low fraction of defects (surfaces, interfaces, dislocations) have negligible influence on phase separation. This is not the case in AgCu nanoalloys where the defect fraction is relatively higher. For instance, dilute amounts of Cu segregate at the sub-surface dislocation cores in AgCu octahedral nanoalloys \cite{settem2021AgCuPCCP} and are energetically favoured in comparison to defect-free segregated arrangement or solid solution } 

For each combination of composition and temperature, we show in Figure \ref{fgr:phase_diagram}a the most probable structure obtained in PTMD. In addition, in the bottom line, we also show the global minimum, which represents the 0 K structures obtained by BHMC. In contrast to the bulk phase diagram, in which only two crystalline phases exist (Ag-rich and Cu-rich solid solutions), nanoalloys exhibit a higher number of structural classes due to finite size and surface effects. We first look at the 0 K structures. \emph{fcc} is observed near pure Ag or pure Cu compositions (n\textsubscript{Cu} $=$ 0 and 33 $-$ 38). As the composition moves away from pure Ag or pure Cu, \emph{Ih (aM)} begins to appear (n\textsubscript{Cu} $=$ 1 $-$ 3 and  24 $-$ 32). The polyicosahedra structures \emph{pIh{$_6$}} and \emph{pIh} are observed at n\textsubscript{Cu} $=$ 4 $-$ 6 and n\textsubscript{Cu} $=$ 8 $-$ 17, respectively. Finally, \emph{7-atom fcc core, chiral shell} is observed at the remaining compositions (n\textsubscript{Cu} $=$ 7 and 18 $-$ 23). Perfect core-shell structures (pure Cu core and pure Ag shell) are observed at n\textsubscript{Cu} $=$ 7 and 8 where the excess energy is among the lowest (see Figure \ref{fgr:gm_excess_energy_structs}). The other structures start appearing only at finite temperatures.

\ms{We now look at the general features of the structural chart at finite temperatures. In most cases, the structures at 0 K are also dominant at finite temperatures with few exceptions close to the boundaries of the different structural regions. Similar to 0 K, \emph{fcc} is observed at the two ends of the composition range with a broader region close to pure Cu. With the addition of few Cu atoms (n\textsubscript{Cu} $=$ 1 $-$ 3) to pure Ag, the dominant structure becomes \emph{Ih (aM)} and at n\textsubscript{Cu} $=$ 3, \emph{Ih (aM)} transforms to \emph{Ih (chiral)} at higher temperatures before becoming disordered. At n\textsubscript{Cu} $=$ 4 $-$ 6, \emph{pIh$_6$} transforms to \emph{\ag{defective} pIh} at higher temperatures. There is a significant composition range (n\textsubscript{Cu} $=$ 7 $-$ 19) where \emph{pIh} is dominant with some of these composition exhibiting \emph{\ag{defective} pIh} at higher temperatures. With increasing Cu content, we observe regions of \emph{7-atom fcc core, chiral shell} and \emph{Ih (chiral)}, in that order. \emph{Ih (chiral)} is the only non-defective structural class that is observed only at higher temperatures as a result of transformation from the low temperature \emph{Ih (aM)}. This will be discussed in more detail in the next section. All the defective structural classes are observed at higher temperatures just below disordered region. A striking feature is the significantly increased thermal stability in the central composition range (\emph{pIh}) compared to pure Cu and pure Ag. Interestingly, the highest thermal stability is not found at the perfect core-shell arrangements (n\textsubscript{Cu} $=$ 7,8 where the excess energy is the lowest), but marginally to the right at n\textsubscript{Cu} $=$ 12,13 (as pointed by the black arrow in Figure \ref{fgr:phase_diagram}a). The structural chart gives the dominant motif, but does not indicate the degree of competition with the other motifs. In Figure \ref{fgr:phase_diagram}b, we plot the Shannon entropy (defined as $\sum_{i} p_iln(p_i)$, where $p_i$ is the probability of each structure class) at a given combination of composition and temperature. A value of zero indicates complete dominance ($p_i=1$) of a given motif and higher values indicate competition with other motifs. We find a greater degree of dominance for \emph{fcc} (close to pure Ag and Cu), \emph{pIh$_6$}, \emph{pIh}, and to some extent for \emph{7-atom fcc core, chiral shell}. Further, the higher thermal stability in the central composition regions of \emph{pIh} is reinforced from the Shannon entropy plot. }


Due to the manner in which  the structural chart of nanoalloys is constructed, it is not straightforward to compare with the bulk phase diagram. The structural chart gives information about the most probable structural class at a given combination of composition and temperature. In contrast, the bulk phase diagram has regions with a single phase and regions where two phases co-exist. In the case of Ag-Cu bulk phase diagram, there are three phases: liquid, Ag-rich $fcc$ solid solution, and Cu-rich $fcc$ solid solution. Below the eutectic temperature, there is a wide region with coexistence of Ag-rich and Cu-rich phases. Thermal stability decreases in the central composition regions and is lowest at the composition X\textsubscript{Cu} $=$ 0.399 corresponding to the eutectic temperature 1052 K \cite{pdCuAg}. On the contrary, for 38-atom nanoalloys, the structural chart indicates increased thermal stability in the central composition region (roughly in the range n\textsubscript{Cu} $=$ 5 to 20, corresponding to copper fractions X\textsubscript{Cu} from 0.132 to 0.526) compared to pure Ag and Cu clusters. This difference stems predominantly from the surface effects in finite-sized particles. A core-shell chemical arrangement with Ag-rich shell covering a pure-Cu core results in relatively low excess energy (see the excess energy curve of global minimum structures in Figure \ref{fgr:gm_excess_energy_structs}), because of the high cohesive energy of Cu combined with low surface energy of Ag. The nanoalloy phase diagrams calculated using classical thermodynamics models with corrections to account for the finite size have the same qualitative features of the bulk phase diagram, with the eutectic shifted at lower temperatures and copper fractions \ms{(ca. 900 K and X\textsubscript{Cu} $\approx$ 0.3 for 4~nm nanoparticles \cite{jabbareh2018})}. This is in contrast to the increased thermal stability shown for compositions around X\textsubscript{Cu} $=$ 0.32 in Figure~\ref{fgr:phase_diagram}a. This discrepancy is rooted in the generic surface corrections that mask the specific structural and chemical configurations that occur in nanoparticles and that cause, for instance, the increased thermal stability of core-shell chemical arrangements.

\begin{figure}[hb]
\centering
  \includegraphics[width=0.5\textwidth]{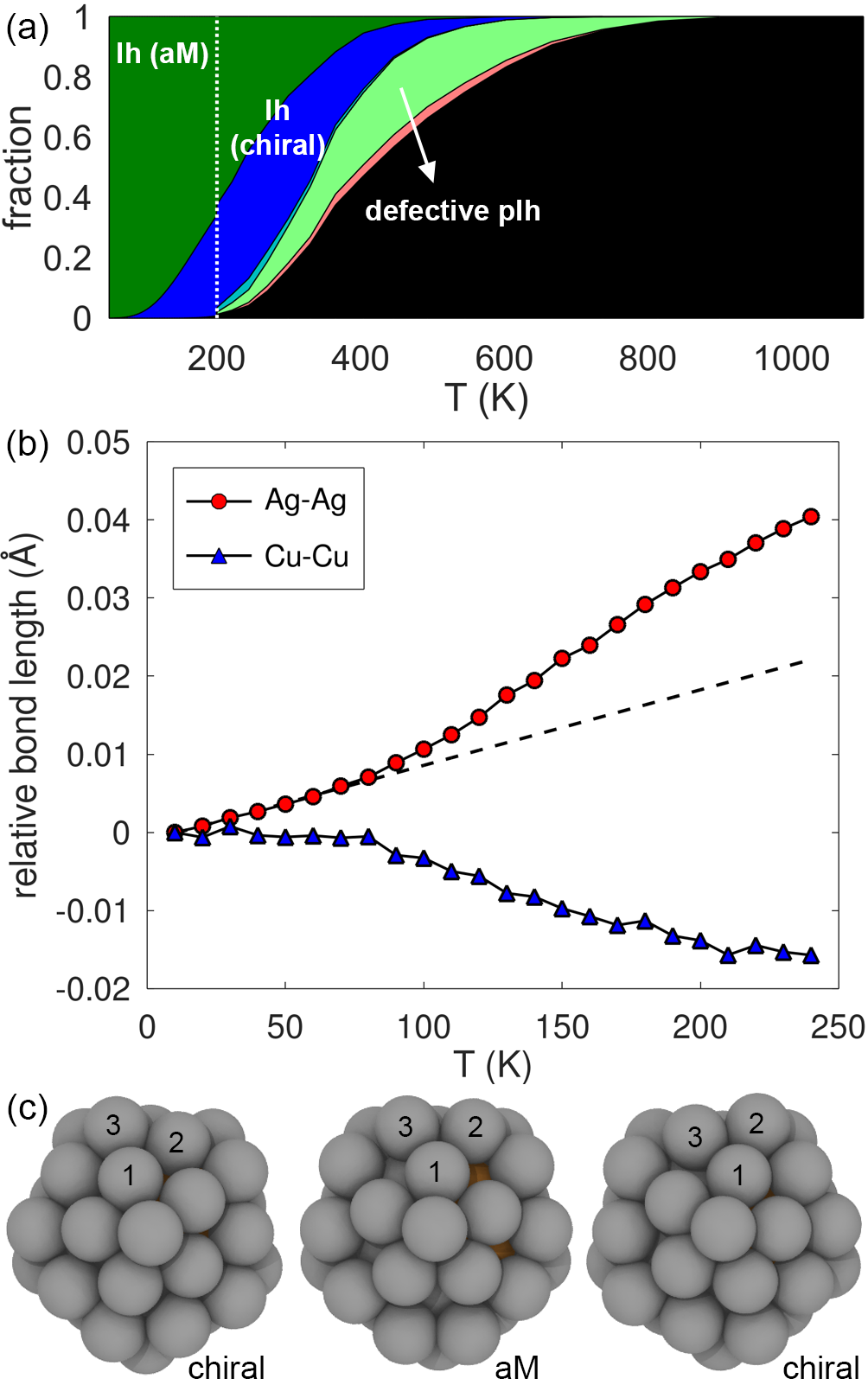}
  \caption{(a) Structural distribution of Ag$_{35}$Cu$_{3}$ nanoalloys in the temperature range 50 K to 1100 K. The dotted line represents the temperature at which calculations from HSA and PTMD are stitched together. (b) Plot of relative bond length of Ag-Ag and Cu-Cu. Dashed line is the linear fit of Ag-Ag bond length. (c) Representative \emph{Ih (aM)} and \emph{Ih (chiral)} arrangements observed at 150 K.}
  \label{fgr:am_to_chiral}
\end{figure}

\subsection{Structural transformations}

\begin{figure}[!b]
\centering
  \includegraphics[width=1.0\textwidth]{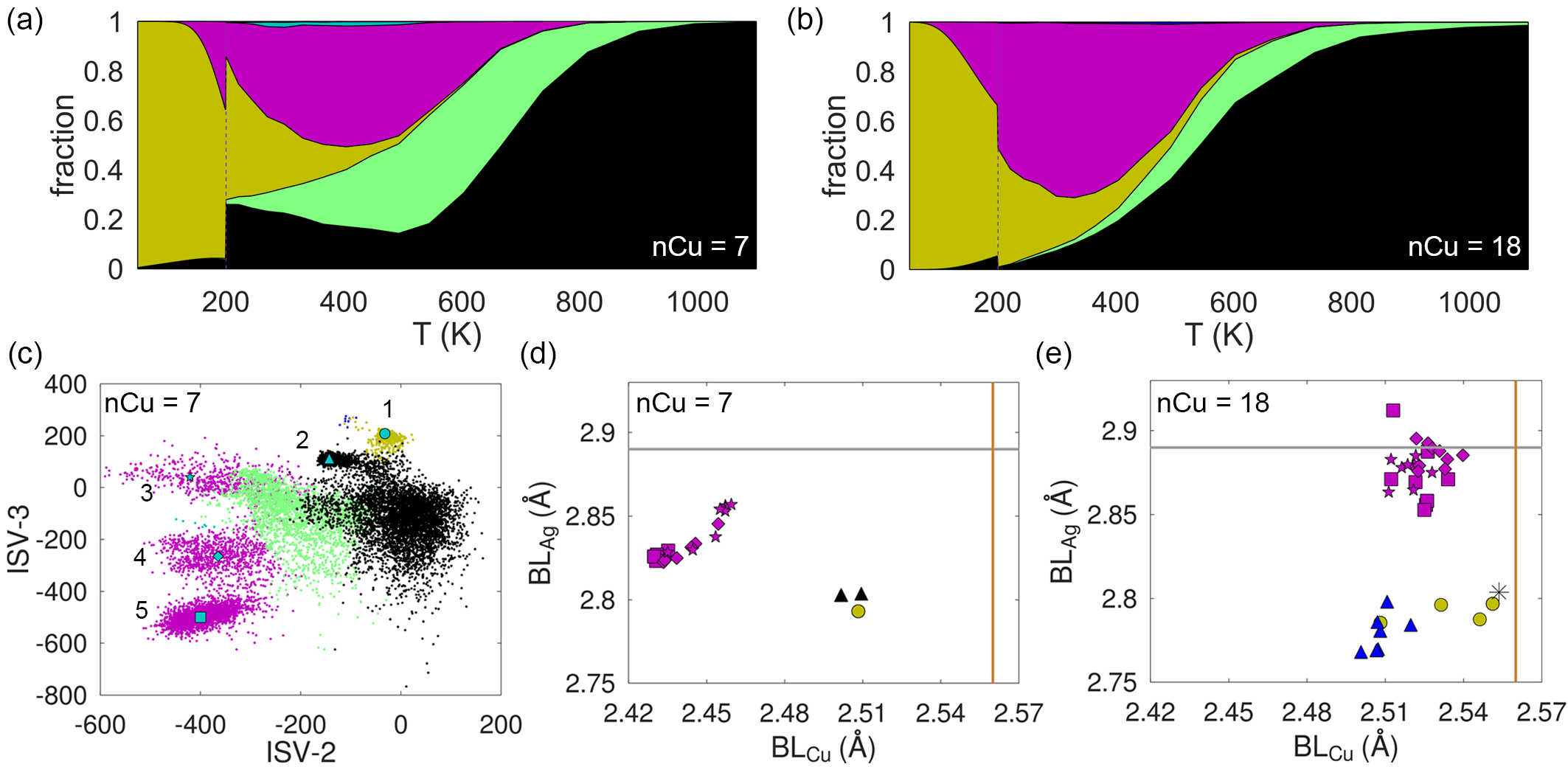}
  \caption{Structural distribution of (a) Ag$_{31}$Cu$_{7}$ and (b) Ag$_{20}$Cu$_{18}$ nanoalloys in the temperature range 50 K to 1100 K. The line at 200 K represents the temperature at which calculations from HSA and PTMD are stitched together. (c) PTMD structures of Ag$_{31}$Cu$_{7}$ projected on the ISV-2/ISV-3 space. \ms{Labels 1 and 2 denote the \emph{7-atom fcc core, chiral shell} and low-temperature disordered structure basins, respectively. Labels 3 to 5 denote the various \emph{pIh} basins.} Mean Cu-Cu (BL\textsubscript{Cu}) and mean Ag-Ag (BL\textsubscript{Ag}) bond lengths of few configurations belonging to various structural classes of  Ag$_{31}$Cu$_{7}$ (d) and  Ag$_{20}$Cu$_{18}$ (e) nanoalloys. The vertical and the horizontal lines correspond to bulk Cu-Cu and Ag-Ag bond length, respectively. Color coding indicates the structural class (same as Figure \ref{fgr:phase_diagram}a). Different symbols indicate the structures from the selected regions marked by \ag{the corresponding cyan symbols} in panel (c) for n\textsubscript{Cu} $=$ 7 and in  Supplementary Figure~\ref{sup_fgr:isv2_isv3_nCu_18} for n\textsubscript{Cu} $=$ 18). The black asterisk in panel (e) refers to the global minimum which has \emph{7-atom fcc core, chiral shell} structure.}
  \label{fgr:stitch_7_atom_nCu_7_18}
\end{figure}

In this section, we look in detail at some of the solid-solid transitions that are observed in this system. These are typically observed at the boundaries where the dominant structural motif changes. Specifically, we focus on \ms{\emph{Ih (aM)} to \emph{Ih (chiral)} transformation which occurs at n\textsubscript{Cu} $=$ 3, 25 $-$ 35} discussing compositions with n\textsubscript{Cu} $=$ 7, 18, 24, 32, and 33. In the following, the structural distribution at selected compositions are estimated by stitching the low-temperature part calculated from HSA (up to 200 K) with the high-temperature part from PTMD (200 K to 1100 K). Some disagreement is to be expected \cite{settem2023AgCuAu} since the harmonic approximation underlying HSA becomes inaccurate as the temperature increases. We observed that the degree of mismatch varies across compositions. However, this approach gives a clear qualitative picture of the structural transformations. For this reason, we did not try to optimize the HSA calculations or the matching temperature further.

The structural distribution (fraction of various structures) at n\textsubscript{Cu} $=$ 3 in the temperature range 50 K to 1100 K is shown in Figure \ref{fgr:am_to_chiral}a. Barring the minor quantitive mismatch, the fractions from HSA and PTMD are in fair agreement. The transformation from \emph{Ih (aM)} to \emph{Ih (chiral)} begins around 100 K. We carried out heating simulations beginning from the global minimum \emph{aM} structure in the range 10 K up to 500 K raising the temperature in steps of 10 K. Prior studies have shown that the mean Ag-Ag (Cu-Cu) bond length increases (decreases) during the thermal transformation from \emph{aM} to \emph{chiral} \cite{settem2020aMchiral1,settem2020aMchiral2}. Figure \ref{fgr:am_to_chiral}b shows the plot of the mean bond length (averaged over six simulations and measured relative to the mean bond length at 10 K) with temperature. Deviation from the low temperature linear region around 100 K (both Ag-Ag and Cu-Cu) correlates well with the changes observed in the structural distribution at low temperature signalling the beginning of the transformation. Geometrically, this transformation occurs through a concerted rotation (either clockwise or counterclockwise) \cite{bochicchio2010chiralIh} of \{111\}-like surface facets (see the facet 1-2-3 marked in \emph{aM} and \emph{chiral} structures in Figure \ref{fgr:am_to_chiral}c).

Referring to Figure \ref{fgr:phase_diagram}a, on either side of the dominant polyicosahedra region, the \emph{7-atom fcc core, chiral shell} undergoes transformation to \emph{pIh}. The structural distribution at n\textsubscript{Cu} $=$ 7 and 18 (Figures \ref{fgr:stitch_7_atom_nCu_7_18}a, b) reveals this transformation. In both cases, the \emph{7-atom fcc core, chiral shell} dominates at low temperature and gradually transforms to \emph{pIh} and \emph{defective pIh}. In the case of n\textsubscript{Cu} $=$ 7, a significant disordered fraction is observed at lower temperatures which decreases gradually and then further increases. The low-temperature disordered structures are different from those at high temperature, which becomes clear when we look at the position of the various structures in the ISV space. \ms{As previously discussed, ISV-1 correlates roughly with the nanoalloy composition and hence, we analyze the various structures in the ISV-2 ISV-3 projection.}
For n\textsubscript{Cu} $=$ 7, Figure \ref{fgr:stitch_7_atom_nCu_7_18}c  shows that a central \emph{defective pIh} region connects the \emph{7-atom fcc core, chiral shell} and the disordered regions on the right to the various \emph{pIh} regions on the left. The low-temperature disordered region (marked as 2) is clearly separated from the large disordered region at high temperatures. These disordered structures are closely related to the \emph{7-atom fcc core, chiral shell} structures. The 7-atom Cu core undergoes a rearrangement (see Supplementary Figure~\ref{sup_fgr:disordered_structs_nCu_7}) where one of the vertex atom relocates to a \{111\} facet. This results in a twisted core and hence the disordering.

In order to understand why the 7-atom fcc core structures become favorable in comparison to icosahedral structures at low temperature, we look at the mean Ag-Ag (BL\textsubscript{Ag}) and Cu-Cu (BL\textsubscript{Cu}) bond lengths (Figures \ref{fgr:stitch_7_atom_nCu_7_18}d, e) of few representative structures (quenched to 0 K) in the regions marked by cyan markers in Figure~\ref{fgr:stitch_7_atom_nCu_7_18}c for n\textsubscript{Cu} $=$ 7 and Supplementary Figure~\ref{sup_fgr:isv2_isv3_nCu_18} for n\textsubscript{Cu} $=$ 18. The vertical and horizontal lines indicate the bond lengths of bulk Cu and Ag, respectively. Structures close to the intersection of the lines result in optimal bond lengths. 7-atom fcc core structures result in favorable Cu-Cu bonds while \emph{pIh} have favorable Ag-Ag bonds. At higher amounts of Cu (n\textsubscript{Cu} $=$ 18 to 24), where \emph{7-atom fcc core, chiral shell} is the dominant motif, optimal Cu-Cu bonds are preferred due to Cu being more cohesive than Ag. This analysis provides a rationale as to why this motif is preferred at certain compositions.

\begin{figure}[!h]
\centering
  \includegraphics[width=1.0\textwidth]{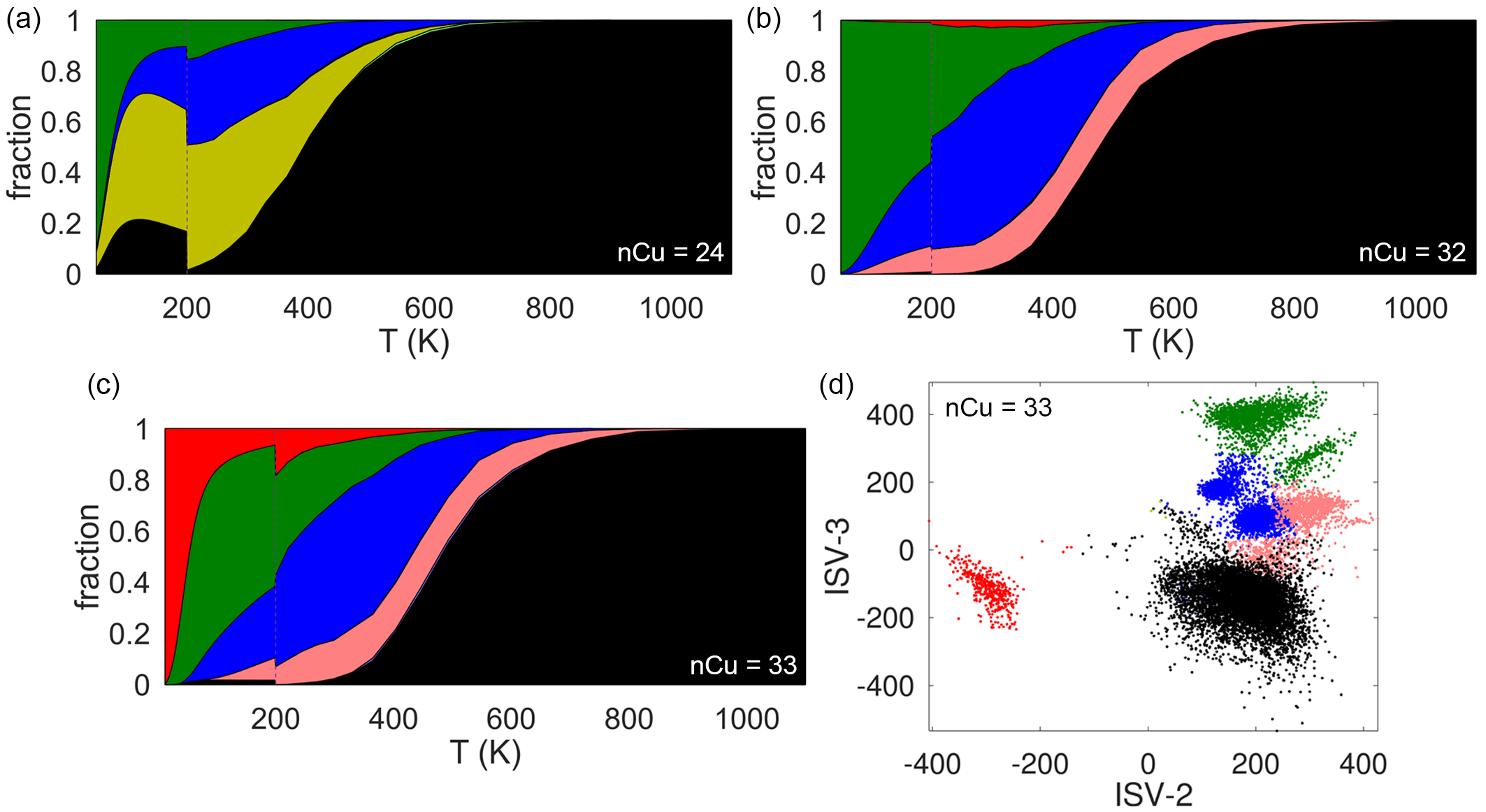}
  \caption{Structural distribution of (a) Ag$_{14}$Cu$_{24}$, (b) Ag$_{6}$Cu$_{32}$, and (c) Ag$_{5}$Cu$_{33}$ nanoalloys in the temperature range 50 K to 1100 K. The line at 200 K represents the temperature at which calculations from HSA and PTMD are stitched together. (d) PTMD structures of Ag$_{5}$Cu$_{33}$ projected on to ISV-2/ISV-3 space.}
  \label{fgr:stitch_nCu_24_32_33}
\end{figure}

Beyond n\textsubscript{Cu} $\approx$ 20, the global minimum changes in the following order: \emph{7-atom fcc core, chiral shell}, \emph{Ih (aM)}, and \emph{fcc}. At n\textsubscript{Cu} $=$ 24, three motifs compete and the structural distribution (Figure \ref{fgr:stitch_nCu_24_32_33}a) shows that the global minimum \emph{Ih (aM)} transforms to \emph{7-atom fcc core, chiral shell} (the dominant motif at finite temperature) and \emph{Ih (chiral)}. As the amount of Cu increases, the proportion of \emph{7-atom fcc core, chiral shell} decreases and \emph{fcc} starts appearing. The structural distribution at n\textsubscript{Cu} $=$ 32 and 33 (Figures \ref{fgr:stitch_nCu_24_32_33}b, c) shows the competition between \emph{Ih (aM)}, \emph{Ih (chiral)}, and \emph{fcc} with \emph{Ih (chiral)} being the dominant motif at higher temperatures. The global minimum \emph{fcc} (at n\textsubscript{Cu} $=$ 33), transforms at low temperature ($<$ 100 K) to a dominant \emph{Ih (aM)} and at higher temperatures to \emph{Ih (chiral)}. The ISV-2/ISV-3 projection of n\textsubscript{Cu} $=$ 33 shows that the \emph{fcc} structural family is rather isolated compared to the other structures. This hints at an interesting structural pathway to \emph{Ih (aM)} which is the dominant motif at low temperatures. \emph{fcc} would go through disordered structures to \emph{Ih (chiral)}, and finally to \emph{Ih (aM)}, i.e., there is no direct pathway from \emph{fcc} to \emph{Ih (aM)}. We defer a detailed study of the geometrical changes and the associated structural transformation pathways to future work.


\section{Conclusions}
\label{sec:conclusions}
In this work, we have generated a finite-temperature structural chart of silver-copper nanoalloys with 38 atoms considering all compositions. First, the equilibrium configurations were sampled using PTMD which results in a huge number of configurations (ca. 2.8 million). We found that the radial distribution function of a given atomic configuration encodes the local atomic arrangement, the global shape of the cluster, and the chemical composition (due to the different bond lengths of atomic pairs). Using RDF as input to a convolutional autoencoder, these configurations were then automatically mapped to a 3-dimensional ISV space according to the approach described in Ref.~\cite{telari2025isv}. In this convenient low-dimensional representation it was possible to classify the various structural families belonging to \emph{octahedra}, \emph{icosahedra}, or a combination of the two geometries (\emph{7-atom fcc core, chiral shell}). This allowed us to construct the structural chart that identifies the most probable structure at a specific composition and temperature. \emph{fcc} structures are found close to pure Cu and Ag clusters. As one moves away from the elemental clusters, icosahedra based structures appear with pockets of compositions where \emph{7-atom fcc, chiral shell} is observed. Typically, low temperature structures are also dominant at higher temperatures with the exception of \emph{Ih (aM)} which transforms to \emph{Ih (chiral)}  and of the emergence of distorted structures at higher temperatures. Close to the boundaries of the various structural regions in the chart, we observe solid-solid transitions. The structural chart highlights the increased thermal stability  at compositions near core-shell chemical ordering which is found at intermediate compositions, \ms{n\textsubscript{Cu} $=$ 5 to 20}. 
This result show that it is essential to incorporate chemical-ordering information into methods that try to adapt bulk phase diagram calculations to finite sized particles, which otherwise reproduce the bulk behavior of decreased thermal stability close to the eutectic composition. Building on this work, the next step would be to study solid solution solubility by considering relatively larger sized nanoalloys in order to address whether solubility is enhanced at the nanoscale. We are currently developing a method capable of efficiently sampling the structural diversity and also the chemical ordering which is challenging for larger nanoalloys.

In summary, we have demonstrated a computational framework that allows to compute, classify, and chart the wealth of structures and compositions of nanoalloys. Computing this structural chart was enabled by a combination of: (i) advanced sampling of the equilibrium structures, (ii) machine-learning classification of the structures, and (iii) calculation of the structural chart as a function of temperature and composition. At the core of our approach is a convolutional neural network that could automatically learn a single 3-dimensional representation that maps structures across temperatures and compositions. As previously shown for elemental nanoclusters \cite{telari2025isv}, this low-dimensional embedding opens the door, now also in the field of nanoalloys, to biased simulation and to the interpretation of experiments and non-equilibrium simulations. Although we demonstrated this approach for a model system of 38-atom AgCu, this framework is completely general and can be used to study binary and multicomponent nanoalloys of any size.

\bibliographystyle{unsrt}  
\bibliography{references}





\clearpage

\section*{Supplementary Information}

\newcommand{\figprefix}{S}
\newcommand{\tabprefix}{S}
\newcommand{\secprefix}{S}

\renewcommand{\thefigure}{\figprefix\arabic{figure}}
\renewcommand{\thetable}{\tabprefix\arabic{table}}
\renewcommand{\thesection}{\secprefix\arabic{section}}
\renewcommand{\thesubsection}{\thesection.\arabic{subsection}}

\setcounter{figure}{0}
\setcounter{table}{0}
\setcounter{section}{0}
\setcounter{subsection}{0}

\begin{table*}[!h]
\small
\caption{Replica temperatures used for PTMD}
\centering
{\def\arraystretch{1.25}
\begin{tabular*}{0.25\textwidth}{@{\extracolsep{\fill}}cc}
\hline
{Replica \#} & {Temperature (K)} \\
\hline
1 & 200 \\
2 & 221 \\
3 & 244 \\
4 & 270 \\
5 & 299 \\
6 & 330 \\
7 & 365 \\
8 & 404 \\
9 & 446 \\
10 & 493 \\
11 & 545 \\
12 & 603 \\
13 & 666 \\
14 & 737 \\
15 & 814 \\
16 & 900 \\
17 & 995 \\
18 & 1100 \\
\hline
\end{tabular*}
}
\label{sup_tab:replica_T}
\end{table*}

\section{Autoencoder architecture and training}
\label{sup_sec:AE}
Here we describe the autoencoder (AE) used to generate the ISV space. The AE is built using PyTorch library \cite{paszke2017automatic}. The neural network works by taking as inputs the RDF of a given non-minimized configuration and giving back as output the RDF of the corresponding minimized configuration.

The AE consists of an encoder and a decoder, corresponding to the fist and second halves of the network, respectively. The encoder is composed by 5 convolutional layers with an increasing number of channels (8, 16, 32, 64, 128),  all having a kernel sizes of 5. Each layers is followed by a rectified linear unit (ReLU) activation, a batch normalization layer and a max-pooling layer, with a kernel size and striding of 2. The decoder mirrors this architecture, replacing convolutional layers with deconvolutional layers, and performing upsampling in place of the encoder max-pooling. At the bottleneck, after the last layer of the encoder, the encoder output is flattened and passed through three fully-connected layers, being the middle one the bottleneck which sets the severity of the dimensionality reduction. Two convolutional layers are used as input and output layers.  

During the training of AE, the input RDF features were normalized in the range [0, 1] across all the dataset. Same normalization was applied to the target RDFs. The training dataset was composed by 561600 RDFs of non-minimized structures as input data ((1/5\textsuperscript{th} of PTMD data)), paired with an equal number of RDFs of the corresponding minimized configurations as target data. The dataset was split into training and validation sets, with 20\% used for validation. During the training, configurations were processed in batches of 64. The AE was trained using MSE loss function with Adam optimizer \cite{AdamOpt}. The starting learning rate was set to 0.005 and then updated using a step scheduler halving its value at epoch number 75 and 125. Training was performed for a total of 150 epochs. In Figure \ref{fig:torchsummary}, we show the summary of the AE architecture printed by torchsummary.

\begin{figure}[h!]
\centering
\begin{lstlisting}[basicstyle=\ttfamily\tiny]
    ----------------------------------------------------------------
        Layer (type)               Output Shape         Param #
================================================================
            Conv1d-1               [-1, 8, 200]              48
         MaxPool1d-2               [-1, 8, 100]               0
              ReLU-3               [-1, 8, 100]               0
       BatchNorm1d-4               [-1, 8, 100]              16
            Conv1d-5              [-1, 16, 100]             656
         MaxPool1d-6               [-1, 16, 50]               0
              ReLU-7               [-1, 16, 50]               0
       BatchNorm1d-8               [-1, 16, 50]              32
            Conv1d-9               [-1, 32, 50]           2,592
        MaxPool1d-10               [-1, 32, 25]               0
             ReLU-11               [-1, 32, 25]               0
      BatchNorm1d-12               [-1, 32, 25]              64
           Conv1d-13               [-1, 64, 25]          10,304
        MaxPool1d-14               [-1, 64, 12]               0
             ReLU-15               [-1, 64, 12]               0
      BatchNorm1d-16               [-1, 64, 12]             128
           Conv1d-17               [-1, 64, 12]          20,544
        MaxPool1d-18                [-1, 64, 6]               0
             ReLU-19                [-1, 64, 6]               0
      BatchNorm1d-20                [-1, 64, 6]             128
           Conv1d-21               [-1, 128, 6]          41,088
        MaxPool1d-22               [-1, 128, 3]               0
             ReLU-23               [-1, 128, 3]               0
      BatchNorm1d-24               [-1, 128, 3]             256
           Conv1d-25               [-1, 128, 3]          82,048
        MaxPool1d-26               [-1, 128, 1]               0
             ReLU-27               [-1, 128, 1]               0
      BatchNorm1d-28               [-1, 128, 1]             256
          Flatten-29                  [-1, 128]               0
           Linear-30                    [-1, 3]             387
           Linear-31                  [-1, 128]             512
             ReLU-32                  [-1, 128]               0
         Upsample-33               [-1, 128, 2]               0
  ConvTranspose1d-34               [-1, 128, 2]          82,048
             ReLU-35               [-1, 128, 2]               0
      BatchNorm1d-36               [-1, 128, 2]             256
         Upsample-37               [-1, 128, 4]               0
  ConvTranspose1d-38                [-1, 64, 6]          41,024
             ReLU-39                [-1, 64, 6]               0
      BatchNorm1d-40                [-1, 64, 6]             128
         Upsample-41               [-1, 64, 12]               0
  ConvTranspose1d-42               [-1, 64, 12]          20,544
             ReLU-43               [-1, 64, 12]               0
      BatchNorm1d-44               [-1, 64, 12]             128
         Upsample-45               [-1, 64, 24]               0
  ConvTranspose1d-46               [-1, 32, 24]          10,272
             ReLU-47               [-1, 32, 24]               0
      BatchNorm1d-48               [-1, 32, 24]              64
         Upsample-49               [-1, 32, 48]               0
  ConvTranspose1d-50               [-1, 16, 50]           2,576
             ReLU-51               [-1, 16, 50]               0
      BatchNorm1d-52               [-1, 16, 50]              32
         Upsample-53              [-1, 16, 100]               0
  ConvTranspose1d-54               [-1, 8, 100]             648
             ReLU-55               [-1, 8, 100]               0
      BatchNorm1d-56               [-1, 8, 100]              16
         Upsample-57               [-1, 8, 200]               0
           Conv1d-58               [-1, 1, 200]              41
================================================================
Total params: 316,836
Trainable params: 316,836
Non-trainable params: 0
----------------------------------------------------------------
Input size (MB): 0.00
Forward/backward pass size (MB): 0.31
Params size (MB): 1.21
Estimated Total Size (MB): 1.52
----------------------------------------------------------------
\end{lstlisting}
\caption{Torchsummary of the autoencoder architecture.}
\label{fig:torchsummary}
\end{figure}


\begin{figure}[!h]
\centering
  \includegraphics[width=0.8\textwidth]{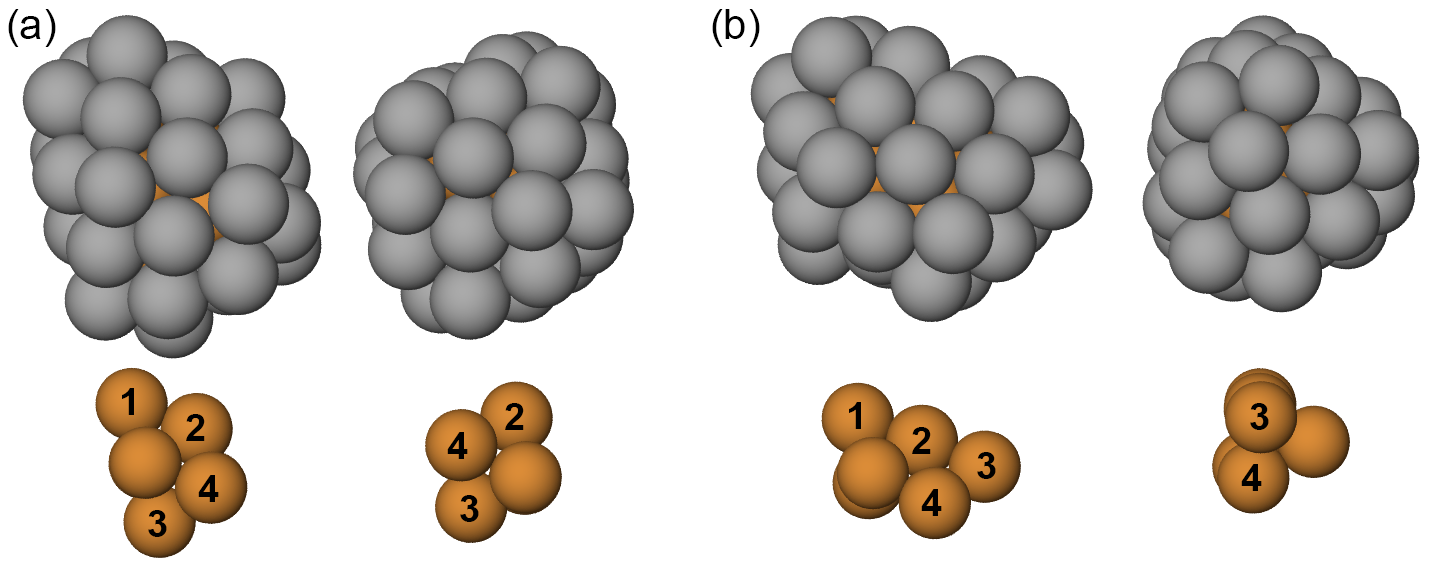}
  \caption{(a) global minimum (7-atom fcc core, chiral shell) and (b) disordered structures of Ag$_{31}$Cu$_{7}$ viewed in two different directions. The bottom images show the 7-atom Cu core. The structure in panel (b) belongs to the disordered region marked in Figure \ref{fgr:stitch_7_atom_nCu_7_18}c in the main manuscript.}
  \label{sup_fgr:disordered_structs_nCu_7}
\end{figure}

\begin{figure}[!h]
\centering
  \includegraphics[width=0.5\textwidth]{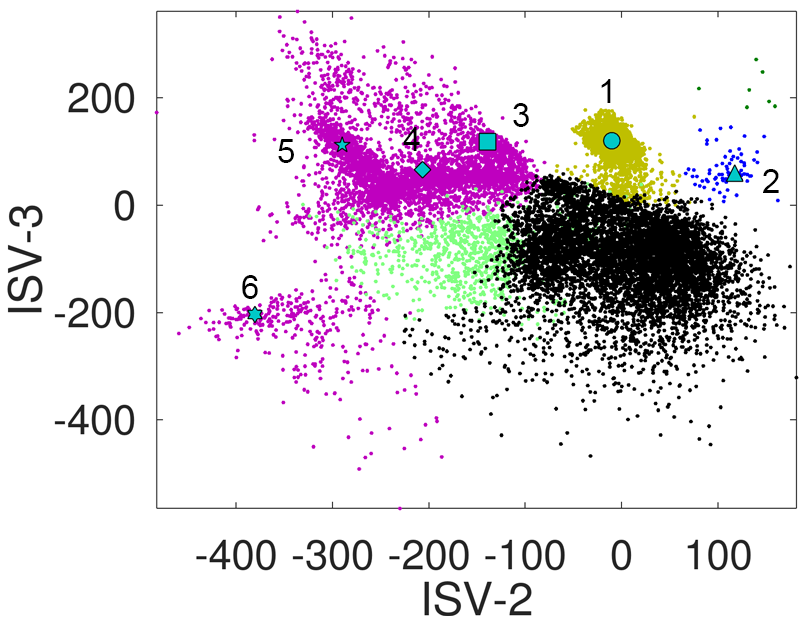}
  \caption{PTMD structures of Ag$_{20}$Cu$_{18}$ projected on to ISV-2 ISV-3 space. \ms{Labels 1 and 2 denote the \emph{7-atom fcc core, chiral shell} and \emph{Ih (chiral)} basins, respectively. Labels 3 to 6 denote the various \emph{pIh} regions, the structures from which are used to calculate the mean bond lengths.} }
  \label{sup_fgr:isv2_isv3_nCu_18}
\end{figure}

\end{document}